\documentclass[a4paper,12pt]{article}

\usepackage{graphics}

\begin{document}

\title{
%\begin{flushright}
%{\normalsize IRB-TH-??/08}
%\end{flushright}
%\vspace{2 cm}
\bf The inhomogeneous equation of state and the speed of convergence to a small cosmological constant}

\author{Hrvoje \v Stefan\v ci\'c\thanks{shrvoje@thphys.irb.hr}
}

\vspace{3 cm}
\date{
\centering
Theoretical Physics Division, Rudjer Bo\v{s}kovi\'{c} Institute, \\
   P.O.Box 180, HR-10002 Zagreb, Croatia}

%\institute{
%  Theoretical Physics Division, Ru\dj er Bo\v{s}kovi\'{c} Institute,
%   P. O. Box 1016, HR-10001 Zagreb, Croatia}

\maketitle

\abstract{The mechanism for the relaxation of the cosmological constant is studied and elaborated. In the model used for the analysis of the relaxation mechanism the universe contains two components: a cosmological constant of an arbitrary size and sign and a component with an inhomogeneous equation of state. Owing to the dynamics of the second component the universe asymptotically tends to a de Sitter phase of expansion characterized by a small effective positive cosmological constant.
An analysis of the asymptotic expansion for a general inhomogeneous equation of state of the second component is made. Several concrete examples are presented and the stability and speed of convergence to their fixed points are analyzed. It is found that the speed of convergence to a fixed point is large whenever the absolute value of the cosmological constant is large.}

\vspace{2cm}

\section{Introduction}

\label{int}

%Imagine that you are walking with an average person at night blessed by sky
%without clouds in an area still not excessively light polluted. If the said person happens to know that you are a physicist/astrophysicist/cosmologist, he/she might ask for a simple description of the universe. Actually, it is even not impossible that the same person might ask you to describe the most important property of the present universe in a single word.
%Then, after some thinking watching the twinkling stars, not very few of us would say: "Well, as much as we know at present, the universe is largely dark". The other person might take it as a joke, might interpret it as a clear absence of your communication skills or even get a bit angry thinking: "That much I could see myself". 

The understanding of the structure, dynamics and composition at cosmological scales is one of the biggest challenges for physics and maybe even for science in general. The results of cosmological observations in recent years \cite{SN,WMAP,LSS} and numerous attempts of their theoretical explanation \cite{Rev} have lead us to an astonishing picture of the present universe. We presently understand only a minor part of the composition of the universe whereas its unknown part is attributed to the cosmic dark sector: dark matter and dark energy. Additional peculiar feature of the universe at the present epoch is its accelerated expansion. The dynamics or mechanism leading to the accelerated expansion became the hot topic in cosmology soon after the accelerated expansion had been established observationally . Almost every element of our, until then valid, picture of the universe and its dynamics, came under scrutiny in search of the acceleration mechanism that would be consistent with the observational data. 

First attempts intervened into the composition of the universe: the existence of a new component with negative pressure (or even several of them) was postulated. The proposals for the nature of this new component, called {\em dark energy}, are numerous \cite{Rev}. To name just a few, dark energy models range from the dynamics of scalar field in the potential \cite{scalar}, which further comprises quintessence, k-essence and phantom energy models, various cosmic barotropic fluid models \cite{bar} to the very concept of static or dynamic cosmological term \cite{dynCC}. Further attempts towards the acceleration mechanism questioned general relativity as a theory of gravitation at cosmological scales \cite{modgrav}. The very concept of our universe as four-dimensional was also relaxed in braneworld models in the quest for the explanation of acceleration \cite{brane}. The list of ingenious approaches to this problem certainly does not end here.

The problem of the cosmological constant (CC) \cite{Wein,Stra,Nob} enjoys a special place among the multitude of proposals for the acceleration of the universe. The $\Lambda$CDM model is probably the simplest and the most popular model of the dark sector. The CC problem is an old one and it comes in many forms. From the viewpoint of fundamental theories, such as quantum field theory (QFT), the size of $\Lambda$ consistent with observations is in a notoriously huge discrepancy with the predictions of these theories. This is a so called ``old CC problem". An additional problem comes from the fact that the present energy densities of matter and CC are of the same order of magnitude, despite their significantly different scaling with the expansion of the universe. This is a so called ``coincidence problem". Although the $\Lambda$CDM model generally fits the data very well, other challenges to $\Lambda$CDM cosmology have been identified \cite{Peri}. However, the importance of the CC problem and its possible resolution is even bigger given that it underlies many other models of the cosmic acceleration, such as dynamical dark energy models.

In this paper we further elaborate the model of relaxation of the cosmological constant introduced in \cite{ccrelax} and studied in \cite{ccrelaxb}. The relaxation of the cosmological constant is understood as a dynamical process in which the universe with a large CC asymptotically tends to a de Sitter regime. The Hubble constant $H^2$ characterizing this de Sitter phase is equivalent to a small effective cosmological constant, $H^2=\Lambda_{eff}/3$. We study the stability and convergence properties of the said model and provide several examples.

\section{The model of the cosmological constant relaxation}

In this section we present a short summary of the CC relaxation model introduced in \cite{ccrelax} and discussed in \cite{ccrelaxb}. We consider a two component cosmological model. The first component is a cosmological constant of arbitrary size and sign, whereas the second component is described by an inhomogeneous equation of state (EOS). The expansion of the universe is given by the Friedmann equation 

\begin{equation}
\label{eq:H2}
H^2=\frac{8 \pi G}{3} (\rho_{\Lambda}+\rho) \, .
\end{equation}
The dynamics of the second component is described by the inhomogeneous EOS of the type

\begin{equation}
\label{eq:p}
p=w \rho - 3 \zeta_0 H^{\alpha+1} \, .
\end{equation}

The formalism of the inhomogeneous EOS proves to be very useful in the description of various cosmological phenomena \cite{Odin1,inhom0, inhom1,inhom2,Bamba,Mota,Jamil}.
It has been shown \cite{Odin1} that the inhomogeneous equation of state can be interpreted as an effective description of modified gravity or braneworld dynamics \cite{Odinmodgrav,Faraonimodgrav}. Another possible interpretation of an inhomogeneous EOS related to bulk viscosity \cite{Weinvisc,Zim,Gron,colistete,avelino,avelino2,avelino3} and in particular its generalizations. 

Using (\ref{eq:H2}) and (\ref{eq:p}) with the standard continuity equation for the second component we obtain a dynamical equation for the Hubble function $H^2$:

\begin{equation}
\label{eq:dynH}
d H^2 + 3(1+w)\frac{da}{a} \left( H^2 - \frac{8 \pi G \rho_{\Lambda}}{3} - \frac{8 \pi G \zeta_0}{1+w} (H^2)^{(\alpha+1)/2} \right) =0\, .
\end{equation}
Using the notation

\begin{equation}
\label{eq:notation}
h=(H/H_X)^2, \;\; s=a/a_X, \;\; \lambda=8 \pi G \rho_{\Lambda}/3 H_X^2, \;\;
\xi=8 \pi G \zeta_0 H_X^{\alpha-1}/(1+w)\, , 
\end{equation}
where $H(a_X)=H_X$, we arrive at the equation relating dimensionless quantities 

\begin{equation}
\label{eq:dynH2}
s \frac{d h}{d s} + 3 (1+w) (h - \lambda - \xi h^{(\alpha+1)/2})=0 \, ,
\end{equation}
and the initial condition $h(1)=1$. 

The relaxation mechanism for the cosmological constant is realized in the regime $\alpha<-1$. For an analytically tractable case $\alpha=-3$, it is straightforward to show \cite{ccrelax} that for large absolute values of the rescaled CC term $\lambda$ of both signs it is possible to obtain the relaxation of the cosmological constant.

%\begin{equation}
%\label{eq:alpha-3}
%\frac{h \, d h}{h^2-\lambda h - \xi} = -3 (1+w) \frac{d s}{s} \, .
%\end{equation}

%\begin{equation}
%\label{eq:hstar1}
%h_{*1}= \frac{1}{2} \left(\lambda+\sqrt{\lambda^2+4 \xi}\right)\, , \\
%\end{equation}
%\begin{equation}
%\label{eq:hstar2}
%h_{*2}= \frac{1}{2} \left(\lambda-\sqrt{\lambda^2+4 \xi}\right) \, . 
%\end{equation}

%\begin{equation}
%\label{eq:solrelax}
%\left( \frac{h-h_{*1}}{1-h_{*1}} \right)^{A_{1}} \left( \frac{h-h_{*2}}{1-h_{*2}} \right)^{A_{2}} = s^{-3(1+w)} \, ,
%\end{equation}
%where $A_1=h_{*1}/(h_{*1}-h_{*2})$ and  $A_2=-h_{*2}/(h_{*1}-h_{*2})$. 

%\begin{equation}  
%\label{eq:lamnegs0}
%h \sim (1-h_{*1})^{A_1}(1-h_{*2})^{A_2} s^{-3(1+w)} \, .
%\end{equation}

%\begin{equation}
%\label{eq:lamnegslarge}
%\lim_{s \rightarrow \infty} h = h_{*1} \, .
%\end{equation}

%\begin{equation}
%\label{eq:happrox1}
%h_{*1} \simeq \frac{\xi}{|\lambda|} \, ,
%\end{equation}

\begin{figure}
\centerline{\resizebox{0.7\textwidth}{!}{\includegraphics{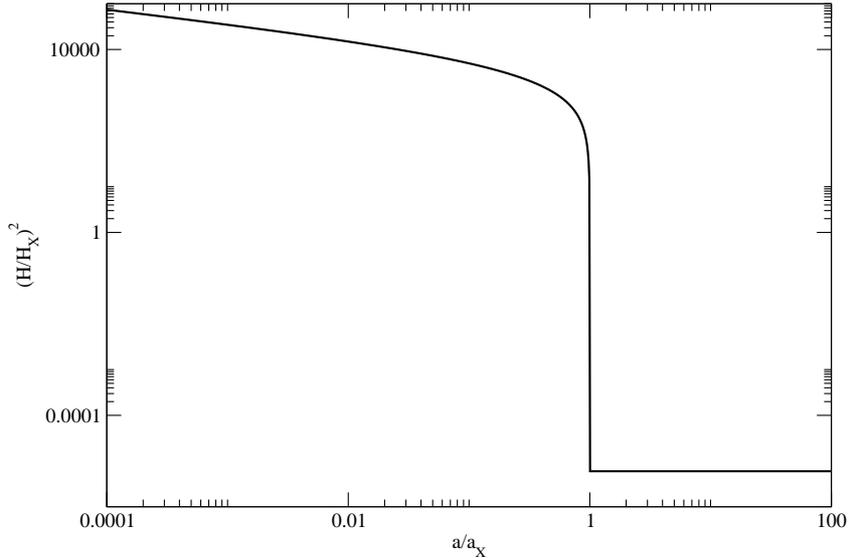}}}
\caption{\label{fig:1} The dynamics of the scaled Hubble function for parameter values $\alpha=-3$, $\lambda=-5000$, $\xi=0.03$ and $w=-0.9$. The dynamics consists of two phases of accelerated expansion connected with an abrupt transition.}
\end{figure}

For negative values of $\lambda$ and $\lambda^2 \gg \xi$ with $\xi>0$ and $w>-1$, the asymptotic expansion of the universe is given by a small value of the Hubble function

\begin{equation}
\label{eq:Hstareq}
\lim_{a \rightarrow \infty} H^2=H_{*1}^2=\frac{24 \pi G \zeta_0}{(1+w) |\Lambda|} \equiv \frac{3 \zeta_0}{(1+w) |\rho_{\Lambda}|}=\frac{\Lambda_{eff}^{-}}{3}\, .
\end{equation}
This small value of $H^2$ corresponds to a small effective positive cosmological constant $\Lambda_{eff}^{-}$. A crucial result is that the $\Lambda_{eff}^{-}$ is small because $\Lambda$ is large in absolute value, as presented in (\ref{eq:Hstareq}). The dynamics of the Hubble function for negative $\lambda$ and some typical values of parameters is given in Fig. \ref{fig:1}.

%\begin{equation}
%\label{eq:lampoys0}
%\lim_{s \rightarrow 0} h = h_{*1} \, 
%\end{equation}

%\begin{equation}
%\label{eq:lampozslarge}
%\lim_{s \rightarrow \infty} h = h_{*2} \, .
%\end{equation}

%\begin{equation}
%\label{eq:happrox2}
%h_{*2} \simeq -\frac{\xi}{\lambda} \, ,
%\end{equation}

\begin{figure}
\centerline{\resizebox{0.7\textwidth}{!}{\includegraphics{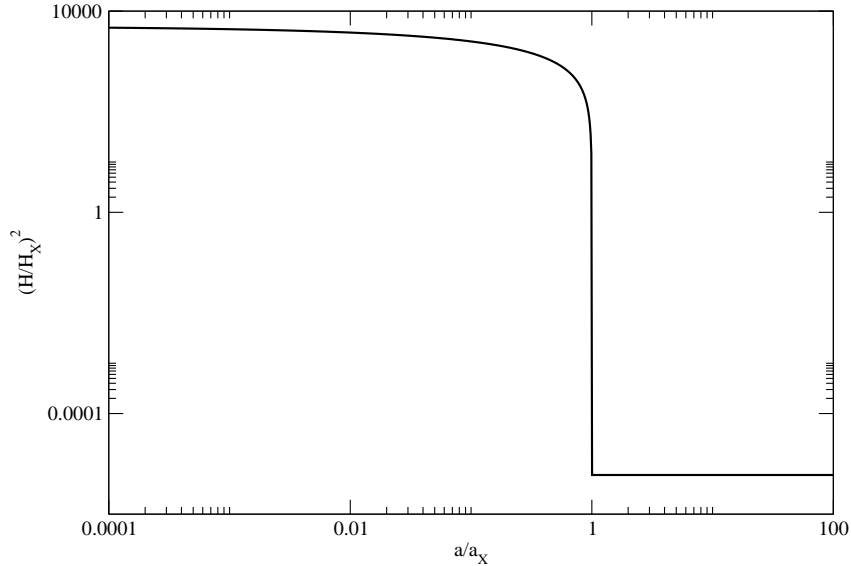}}}
\caption{\label{fig:2} The dependence of the Hubble function on the scale factor for the values of parameters $\alpha=-3$, $\lambda=5000$, $\xi=-0.03$ and $w=-1.1$. Two phases of accelerated expansion are interconnected with an abrupt transition.
}
\end{figure}

For large positive values of $\lambda$ (so that $\lambda^2 \gg |\xi|$) and for $\xi<0$ and $w<-1$, the Hubble function tends asymptotically to  

\begin{equation}
\label{eq:Hstareq2}
\lim_{a \rightarrow \infty} H^2=H_{*2}^2=-\frac{24 \pi G \zeta_0}{(1+w) \Lambda} \equiv -\frac{3 \zeta_0}{(1+w) \rho_{\Lambda}} = \frac{\Lambda_{eff}^{+}}{3}\, .
\end{equation}
Again as in the case of negative $\lambda$, the small asymptotic value of the Hubble function $H^2$ can be interpreted as a small effective positive cosmological constant $\Lambda_{eff}^{+}$. The effective CC is small because $\Lambda$ is large. The dependence of the Hubble function on the scale factor for positive $\lambda$ and some typical values of parameters is depicted in Fig. \ref{fig:2}.

As described above, a simple model defined by (\ref{eq:H2}) and (\ref{eq:p}) provides parameter regimes in which the relaxation of the cosmological constant can be realized. It is important to stress that in this model there is no fine-tuning and the asymptotic de Sitter phase is characterized by an effective positive CC which is small because the size of the real CC $|\Lambda|$ is large. Based on the calculations from e.g. QFT  we could say that the size of $\lambda$ is ``naturally large". Additionally, since $\xi$ describes the deviations from GR or the generalized effects of viscosity we could say that $\xi$ is ``naturally small". Then, from the behavior of the model for $\alpha<-1$ \cite{ccrelax} and in particular from (\ref{eq:Hstareq}) and (\ref{eq:Hstareq2}) we can expect that the size of the effective CC in the asymptotic de Sitter phase is ``naturally small".
This fact corresponds to the resolution of the old CC problem in the studied model.

\section{Stability and convergence analysis}

\label{stability}

Given the potential of the relaxation mechanism described in the preceding sections for the resolution of the cosmological constant problem, it is important to study more general inhomogeneous equations of state for which the relaxation mechanism can be realized. Furthermore, as the dynamics of the Hubble function in the model \cite{ccrelax} exhibits abrupt transition followed by a swift stabilization at the asymptotic value, additional insight into the dynamical details of the approach to the asymptotic value of $H$ is needed. To this end, we consider a general  inhomogeneous equation of state  

\begin{equation}
\label{eq:genin}
p=w \rho - g(H^2) \, ,
\end{equation}
which, using the standard rescaling as given in (\ref{eq:notation}), results in the following dynamical equation for the Hubble parameter:

\begin{equation}
\label{eq:genh}
s \frac{dh}{ds} + 3 (1+w)(h-\lambda-f(h))=0 \, ,
\end{equation}
with the initial condition $h(1)=1$. Here we have $f(h)=((8 \pi G)/(3(1+w)H_X^2)) g(H_X^2 h)$. We further introduce the notation $F(h)= 3 (1+w)(h-\lambda-f(h))$ and $F'(h)=3(1+w)(1-f'(h))$. 

For a relaxation mechanism to be effective, we expect the dynamical system (\ref{eq:genh}) to have a fixed point $h_{*}$ which is much smaller than $|\lambda|$, i.e. for which we have $h_{*} \ll |\lambda|$. From the condition $F(h_{*})=0$ and the said expectation of the size of $h_{*}$ it is straightforward to obtain

\begin{equation}
\label{eq:hstar}
h_{*}=f^{-1}(-\lambda) \, .
\end{equation}   

The sign of the quantity $F'(h_{*})$ determines the stability of the fixed point $h_{*}$, whereas its size controls the speed of convergence to the fixed point. The fixed point $h_{*}$ is stable for $F'(h_{*})>0$ and it is given by the expression
 
\begin{equation}
\label{eq:Fhstar}
F'(h_{*})=3(1+w)(1-f'(f^{-1}(-\lambda))=3(1+w) \left( 1-\frac{1}{(f^{-1})'(-\lambda)} \right) \, .
\end{equation}

We further apply the general analysis displayed above to specific examples of the function $f(h)$. First we turn to the choice $f(h)=\xi/h$, already analyzed in detail in \cite{ccrelax} and presented in the preceding sections of this paper. It is easy to see that in this case we have
\begin{equation}
\label{eq:ex1hstar}
h_{*}=-\frac{\xi}{\lambda} 
\end{equation}
and 

\begin{equation}
\label{eq:ex1Fprime}
F'(h)=3(1+w)\left( 1+\frac{\xi}{h^2} \right) \, .
\end{equation}
 
Combining (\ref{eq:ex1hstar}) and (\ref{eq:ex1Fprime}) and using $\lambda^2 \gg |\xi|$, we obtain 

\begin{equation}
\label{eq:ex1stab}
F'(h_{*}) = 3(1+w) \frac{\lambda^2}{\xi} \, .
\end{equation}
This results sheds additional light on the results obtained in \cite{ccrelax} and \cite{ccrelaxb}. From (\ref{eq:ex1hstar}) and (\ref{eq:ex1stab}) it becomes clear that to have a stable fixed point for $\lambda>0$ we must have $\xi<0$ and $w<-1$, whereas for $\lambda<0$ we need $\xi>0$ and $w>-1$. However, the most important feature of (\ref{eq:ex1stab}) is that the speed of convergence towards the fixed point  $h_{*}$ depends quadratically on $\lambda$. This finding explains why the fixed points in Figs \ref{fig:1} and \ref{fig:2} are reached so swiftly after the transition. 

In the following example we consider the function $f(h)=A_1 \ln h$. For this functional form of $f(h)$ we have
\begin{equation}
\label{eq:ex2hstar}
h_{*}=e^{-\lambda/A_1} 
\end{equation}
and 

\begin{equation}
\label{eq:ex2Fprime}
F'(h_{*})=3(1+w)\left( 1-A_1 e^{\lambda/A_1} \right) \, .
\end{equation}

The condition of the stability of the fixed point $h_{*}$ is $-3(1+w)A_1>0$ (given that the second term in  (\ref{eq:ex2Fprime}) dominates). The speed of convergence to the fixed point grows exponentially with $\lambda$.

The next example employs the functional form $f(h)=A_2 e^{b_1/h}$. As in the preceding examples we have 
\begin{equation}
\label{eq:ex3hstar}
h_{*}=\frac{b_1}{\ln \left(-\frac{\lambda}{A_2} \right)}  
\end{equation}
and 

\begin{equation}
\label{eq:ex3Fprime}
F'(h_{*})=3(1+w)\left(1-\frac{\lambda}{b_1} \left( \ln \left(-\frac{\lambda}{A_2} \right) \right)^2 \right) \, .
\end{equation}
The conditions for the existence of small and positive $h_{*}$ comprise $-\lambda/A_2<0$ and $b_1/\ln (-\lambda/A_2)>0$.
The fixed point $h_{*}$ is stable for $-(1+w)\lambda/b_1<0$. 

Finally, the last example to be considered in this paper is specified by the function $f(h)=A_3 \exp\left( \exp\left( \frac{b_2}{h} \right) \right)$.  We further obtain

\begin{equation}
\label{eq:ex4hstar}
h_{*}=\frac{b_2}{\ln \left(\ln \left(-\frac{\lambda}{A_3} \right) \right)}
\end{equation}
and

\begin{equation}
\label{eq:ex4Fprime}
F'(h_{*})=3(1+w)\left(1-\frac{\lambda}{b_2}  \ln \left(-\frac{\lambda}{A_3} \right) \left( \ln \left(\ln \left(-\frac{\lambda}{A_3} \right) \right) \right)^2 \right) \, .
\end{equation}

The expressions (\ref{eq:ex4hstar}) and (\ref{eq:ex4Fprime}) show the conditions for the existence of a positive fixed point $h_{*}$ are 
$\lambda/A_3 <0$, $\ln(-\lambda/A_3)>0$ and $b_2/\ln (\ln (-\lambda/A_3) )>0$. The fixed point is stable if $ (1+w)\frac{\lambda}{b_2}  \ln \left(-\frac{\lambda}{A_3} \right)<0$. 

The conclusions of the general analysis given at the beginning of  this section and the specific results for the studied examples indicate an interesting characteristic of the speed of convergence to the fixed point $h_{*}$. The faster the growth of the function $f(h)$ when $h$ acquires small values, the smaller the speed of convergence to the fixed point, measured by the dependence of $F'(h_{*})$ on $\lambda$. However, even for the examples with the mildest dependence of $F'(h_{*})$ on $\lambda$, the speed  of convergence is extremely large owing to the fact that $|\lambda|$ is large. The examples discussed in this section were selected primarily to better illustrate this property of the convergence to the fixed point. This is especially true for the last example of double exponential.     

\section{Conclusions}

The results of this paper, along with the findings of \cite{ccrelax} and \cite{ccrelaxb}, show that the mechanism of the CC relaxation based on the inhomogeneous EOS is robust. In section \ref{stability} general conditions for the onset of the CC relaxation mechanism are given. Various inhomogeneous EOS can reproduce the asymptotic de Sitter phase without particular fine-tuning. A general needed feature of the inhomogeneous EOS is that the inhomogeneous term becomes increasingly important as the expansion slows down and its effect finally equilibrate the action of a large CC. Thus the expansion of the universe settles down in a de Sitter phase characterized by a small effective positive cosmological constant. An interesting feature of the models is that the faster the inhomogeneous part, defined by $f(h)$ grows as $h$ becomes small, the slower the approach to the asymptotic value, i.e. fixed point. However, for a large $|\lambda|$, even in the examples with the slowest approach to the fixed point, the speed of convergence is large, principally because of the very size of $|\lambda|$. These results largely explain some features of plots given in Figs \ref{fig:1} and \ref{fig:2}. Namely, given a very large speed of convergence to the fixed points, it is easier to understand the abrupt transition in the dynamics of $h$ and especially an extremely fast approach to the asymptotic value. All these results add to our understanding of the CC relaxation mechanism based on the inhomogeneous EOS and provide a further incentive to incorporate the said mechanism into a complete and workable cosmological model.

{\bf Acknowledgements.} 
The author would like to thank V. Zlati\' c for a valuable comments on the theory of dynamical systems.
This work was supported by the Ministry of Education, Science and Sports of the Republic of Croatia 
under the contract No. 098-0982930-2864.


\begin{thebibliography}{88}
\bibitem{SN} A.G. Riess et al., Astron. J. {\bf 116} (1998) 1009; S. Perlmutter et 
al., Astrophys. J. {\bf 517} (1999) 565;  W. Michael Wood-Vasey et al., Astrophys. J. {\bf 666} (2007) 694; Pierre Astier et al., Astron. Astrophys. {\bf 447} (2006) 31. 
\bibitem{WMAP}  E. Komatsu et al., arXiv:0803.0547 [astro-ph].
\bibitem{LSS} M. Tegmark et al., Astrophys. J. {\bf 606} (2004) 702; M. Tegmark et al., Phys. Rev. D {\bf 69} (2004) 103501.  
\bibitem{Rev} T. Padmanabhan, Phys. Rept. {\bf 380} (2003) 235; E.J. Copeland, M. Sami, S. Tsujikawa, Int. J. Mod. Phys. D {\bf  15} (2006) 1753; J. Frieman, M. Turner, D. Huterer, arXiv:0803.0982 [astro-ph]; T. Padmanabhan, arXiv:0807.2356 [gr-qc].
\bibitem{scalar} 
C. Wetterich, Nucl. Phys. B {\bf 302} (1988) 668;
B. Ratra, P.J.E. Peebles, Phys. Rev. D {\bf 37} (1988) 3406;
R.R. Caldwell, R. Dave, P.J. Steinhardt, Phys. Rev. Lett. {\bf 80} (1998) 1582; 
I. Zlatev, L.-M. Wang, P.J. Steinhardt, Phys. Rev. Lett. {\bf 82} (1999) 896.
S. Nojiri, S.D. Odintsov, Phys. Lett. B {\bf 562} (2003) 147; 
E. Elizalde, S. Nojiri, S.D. Odintsov, Phys. Rev. D {\bf 70} (2004) 043539; 
S. Nojiri, S.D. Odintsov, Phys. Lett. B {\bf 595} (2004) 1; 
S. Nojiri, S.D. Odintsov, S. Tsujikawa, Phys. Rev. D {\bf 71} (2005) 063004; 
S. Nojiri, S.D. Odintsov, Gen. Rel. Grav. {\bf 38} (2006) 1285. 
\bibitem{bar} A.Yu. Kamenshchik, U. Moschella, V. Pasquier, Phys. Lett. B {\bf 511
} (2001) 265; 
N. Bilic, G.B. Tupper, R.D. Viollier, Phys. Lett. B {\bf 535} (2002) 17; 
E.V. Linder, R.J. Scherrer, arXiv:0811.2797 [astro-ph].
\bibitem{dynCC}
I.L. Shapiro, J. Sola, Phys. Lett. B {\bf 475} (2000) 236; 
A. Babic, B. Guberina, R. Horvat, H. Stefancic, Phys. Rev. D {\bf 65} (2002) 085002;
J. Grande, J. Sola, H. Stefancic, JCAP {\bf 0608} (2006) 011;
I.L. Shapiro, J. Sola, arXiv:0808.0315 [hep-th] and references therein.
\bibitem{modgrav} S. Nojiri, S.D. Odintsov, Phys. Rev. D {\bf 68} (2003) 123512; 
S.M. Carroll, V. Duvvuri, M. Trodden, M.S. Turner, Phys. Rev. D {\bf 70} (2004) 043528; 
S. Nojiri, S.D. Odintsov, Gen. Rel. Grav. {\bf 36} (2004) 1765; 
S. Nojiri, S.D. Odintsov, S. Tsujikawa, Phys. Rev. D {\bf 71} (2005) 063004; 
S.M. Carroll, A. De Felice, V. Duvvuri, D.A. Easson, M. Trodden, M.S. Turner, Phys. Rev. D {\bf 71} (2005) 063513; 
S. Nojiri, S.D. Odintsov, Gen. Rel. Grav. {\bf 36} (2004) 1765; 
S. Capozziello, S. Nojiri, S.D. Odintsov, A. Troisi, Phys. Lett. B {\bf 639}  (2006) 135; 
W. Hu, I. Sawicki, Phys. Rev. D {\bf 76} (2007) 064004; 
\bibitem{brane} S. Nojiri, S.D. Odintsov, Phys. Lett. B {\bf 484} (2000) 119;. 
V. Sahni, Yu. Shtanov, JCAP {\bf 0311} (2003) 014; 
S. Nojiri, S.D. Odintsov, Phys. Lett. B {\bf 565} (2003) 1. 
\bibitem{Wein} S. Weinberg, Rev. Mod. Phys. {\bf 61} (1989) 1.
\bibitem{Stra} N. Straumann, in Duplantier, B. (ed.) et al.: Vacuum energy, renormalization, 7-51, 
arXiv:astro-ph/0203330.
\bibitem{Nob} S. Nobbenhuis, arXiv:gr-qc/0609011.
\bibitem{Peri} L. Perivolaropoulos, arXiv:0811.4684 [astro-ph].
\bibitem{ccrelax} H. Stefancic, arXiv:0807.3692 [gr-qc], to appear in Phys. Lett. B.
\bibitem{ccrelaxb} H. Stefancic, arXiv:0811.4548 [gr-qc].
\bibitem{Odin1} S. Nojiri, S.D. Odintsov, Phys. Rev. D {\bf 72} (2005) 023003.
\bibitem{inhom0}  S. Capozziello, V.F. Cardone, E. Elizalde, S. Nojiri, S.D. Odintsov, Phys. Rev. {\bf D73} (2006) 043512. 
\bibitem{inhom1} I. Brevik, E. Elizalde, O. Gorbunova, A.V. Timoshkin, Eur. Phys. J. {\bf C52} (2007) 223.
\bibitem{inhom2} I. Brevik, O.G. Gorbunova, A.V. Timoshkin, Eur. Phys. J. {\bf C51} (2007) 179.
\bibitem{Bamba} K. Bamba, S. Nojiri, S.D. Odintsov, JCAP {\bf 0810} (2008) 045.
\bibitem{Mota} D.F. Mota, C. van de Bruck, Astron. Astrophys. {\bf 421} (2004) 71.
\bibitem{Jamil} M. Jamil, M. Ahmad Rashid, Eur. Phys. J. C {\bf 56} (2008) 429. 
\bibitem{Odinmodgrav} S. Nojiri, S.D. Odintsov, Int. J. Geom. Meth. Mod. Phys. {\bf 4} (2007) 115, hep-th/0601213.
\bibitem{Faraonimodgrav} T.P. Sotiriou, V. Faraoni, arXiv:0805.1726 [gr-qc]. 
\bibitem{Weinvisc} S. Weinberg, Astrophys. J. {\bf 168} (1971) 175.
\bibitem{Zim} W. Zimdahl, Phys. Rev. D {\bf 53} (1996) 5483.
\bibitem{Gron} \O. Gr\o n, Astrophys. Space Sci. {\bf 173} (1990) 191.
\bibitem{colistete} R. Colistete, Jr., J.C. Fabris, J. Tossa, W. Zimdahl, Phys. Rev. {\bf D76} (2007) 103516.
\bibitem{avelino} A. Avelino, U. Nucamendi, F.S. Guzman, AIP Conf. Proc. {\bf 1026} (2008) 300. 
\bibitem{avelino2} A. Avelino, U. Nucamendi, arXiv:0810.0303 [gr-qc].
\bibitem{avelino3} A. Avelino, U. Nucamendi, arXiv:0811.3253 [gr-qc].


\end{thebibliography}
\end{document}